\documentclass[fleqn,twoside]{article}
\usepackage{espcrc2}

\usepackage{psfig}

\newcommand{\AmS}{{\protect\the\textfont2
  A\kern-.1667em\lower.5ex\hbox{M}\kern-.125emS}}

\newcommand{\ba}{\begin{eqnarray}}
\newcommand{\ea}{\end{eqnarray}}

\newcommand{\be}{\begin{equation}}
\newcommand{\ee}{\end{equation}}
\def\ket#1{\vert #1 \rangle}
\def\bra#1{\langle #1 \vert}

\newcommand{\Adag}{{A^{\dag}_\gamma}}
\newcommand{\A}{{A_\gamma}}
\newcommand{\N}{{N_\gamma}}
\newcommand{\n}{{n_\gamma}}

\def\AA{{\cal A}}
\def\AAdag{{\cal A}^{\dag}}
\def\V{\left[ \lambda^2 \sum_n \frac{\bra{0} V \ket{n} \bra{n} V \ket{0}}{(E_n - z)(E_0 - z)} \right]}

\def\Vee{ \lambda^2 \sum_n \frac {\bra{0} V \ket{n} \bra{n} V \ket{0}}{E_n - z} }
\title{\vskip -1.25cm \noindent\hfill\hbox to 1.5in{\rm  } \vskip 1pt \noindent\hfill\hbox
to 1.5in{\rm SLAC-PUB-11532 \hfill  } \vskip 10pt
Adaptive Perturbation Theory: Quantum Mechanics and Field Theory}

\author{Marvin Weinstein\address{Stanford Linear Accelerator Center, Stanford
University \\ Stanford, California 94309 \\USA}%
\thanks{This work was supported by the U.~S.~DOE, Contract
No.~DE-AC03-76SF00515.}}%
\begin{document}

\begin{abstract}
Adaptive perturbation is a new method for perturbatively computing
the eigenvalues and eigenstates of quantum mechanical Hamiltonians
that are widely believed not to be solvable by such methods\cite{Feranchuketal}.
The novel feature of adaptive perturbation theory is that it decomposes
a given Hamiltonian, $H$, into an unperturbed part and a
perturbation in a way which extracts the leading non-perturbative
behavior of the problem exactly.  In this talk I will introduce the
method in the context of the pure anharmonic oscillator and then
apply it to the case of tunneling between symmetric minima. After
that, I will show how this method can be applied to field theory. In
that discussion I will show how one can non-perturbatively extract
the structure of mass, wavefunction and coupling constant
renormalization. \vspace{1pc}
\end{abstract}

\maketitle

\section{Introduction}

To avoid controversy I have titled this talk {\it Adaptive
Perturbation Theory\/}, but I could have equally well called it {\it
Non-Perturbative Perturbation Theory\/} (except that I was told
people would dismiss it as crazy).  Actually I hope you will
agree by the end of the talk that such a title would have been eminently
defensible.

\section{Two Topics}

This talk is divided into two parts.  First I introduce
the general ideas within the framework of ordinary quantum
mechanics. Next, I show how to extend these ideas to the case
of scalar field theory in any number of space-time dimensions.
I will begin the quantum mechanics discussion by talking about the
pure anharmonic oscillator and then extend this discussion to
cover the anharmonic oscillator with either a positive or negative
mass term.

The theory of an anharmonic oscillator with a positive
mass term has a rich literature.  The various authors usually
attempt to resum the ordinary perturbation expansion to obtain the
non-perturbative behavior. I will show that one can avoid such
complication by exhibiting a convergent perturbation expansion for
each eigenstate of the Hamiltonian.  An expansion that, moveover, captures
all non-perturbative features of the problem in zeroth order. The
case of the anharmonic oscillator with a negative mass term will
allow me to extend the simplest adaptive technique to handle the
case of tunneling between symmetric minima.  In this talk I will not
discuss the case of tunneling between very asymmetric minima,
however this topic is covered in the paper to appear on this work.

\section{The Harmonic Oscillator}

Let me begin by reminding you of a some simple facts pertaining to
the ordinary harmonic oscillator.  The Hamiltonian of the ordinary
harmonic oscillator is:
\be
    H = \frac{1}{2m}\, p^2 + \frac{m \omega^2}{2}\, x^2 , \quad {\rm where} \,\,\left[p,x\right] = -i.
\label{harm} \ee
Next, let me introduce $\gamma$-dependent operators $\Adag$ and $\A$ by
\ba
    x &=& {1 \over \sqrt{2 \gamma}}\left( \Adag + \A \right) , \nonumber\\
    p &=& i\,\sqrt{\gamma \over 2} \left( \Adag - \A \right).
\label{annandcr}
\ea
Substituting this into Eq.~\ref{harm} yields
\ba
 H &=& \frac{\gamma}{4\,m} \left( -{\Adag}^2 - \A^2 + 2\,\Adag \A + 1
    \right) \nonumber\\
    &+& \frac{m\,\omega^2}{4\,\gamma} \left( {\Adag}^2 + \A^2 + 2\,\Adag \A + 1
    \right).
\ea
It is customary to choose $\gamma = m\,\omega$ so as to cancel
the terms ${\Adag}^2$ and ${\A}^2$ and render the Hamiltonian diagonal
in the number operator.  This immediately tells us that the eigenstates of the
Hamiltonian are the eigenstates of the number operator and that the energies of
these states are given by $E_n = (n + 1/2)\,\omega$.

\section{The Anharmonic Oscillator}

In the case of the general anharmonic oscillator the Hamiltonian is
\be
    H = \frac{1}{2}\, p^2 + \frac{m\,\omega^2}{2}\,x^2 + \frac{1}{6}\,\lambda\, x^4 .
\label{anharmone}
\ee
The customary way to deal with this problem is two introduce annihilation and
creation operators chosen to diagonalize the first two terms and then construct
a perturbation theory based upon expanding in the coupling $\lambda$.  This
approach has a well known problem; namely, it is known
to diverge, no matter how small $\lambda$, due to $n!$ growth of the
perturbation series. I will now show how to avoid this problem and produce
a rapidly convergent {\it adaptive perturbation theory\/} expansion for each
eigenstate and eigenvalue of $H$.  To emphasize that this is not an expansion
in $\lambda$ I will begin by specializing to the case $m=0$.

Setting $m=0$ in Eq.~\ref{anharmone} and making the substitution for $x$ and $p$ given
in Eq.~\ref{annandcr}, leads to
\ba
    H = \frac{\gamma}{4}&&\hskip-18pt \left( -\Adag^2 -\A^2 + 2\,\Adag \A + 1 \right) \nonumber\\
    + \frac{\lambda}{4\gamma^2}&&\hskip-18pt \left( \Adag^2 \A^2 +  2 \Adag \A + 1
    + \frac{1}{6} (\Adag^4 + \A^4 ) \right.\nonumber\\
    + \frac{2}{3}\left(\right.&&\hskip-18pt \left.\Adag^3 \A + \Adag \A^3 ) + ( \Adag^2 + \A^2 ) \right) \nonumber\\
    = \left( \frac{\gamma}{4}\right.&&\hskip-18pt + \left.\frac{\lambda}{4\gamma^2} \right) ( 2\,\N + 1)
    + \frac{\lambda}{4\gamma^2} \N\,(\N-1) \nonumber\\
    + \left( \frac{\gamma}{4} \right.&&\hskip-18pt \left.- \frac{\lambda}{2\gamma^2} \right) (\Adag^2 + \A^2) \nonumber\\
    + \frac{\lambda}{4\gamma^2} &&\hskip-18pt\left( \frac{1}{6}(\Adag^4 + \A^4)
    + \frac{2}{3} (\Adag^3 \A + \Adag \A^3) \right),\nonumber\\
\ea
where I have defined the $\gamma$-dependent number operator
\be
    \N = \Adag \A .
\ee
Given this expression I define the $\gamma$-dependent unperturbed Hamiltonian, $H_0$ and the perturbation
$V$ by
\ba
    H =&& \hskip-18pt H_0(\gamma) + V(\gamma), \label{pertdecompone} \nonumber\\
    H_0(\gamma)=&&\hskip-18pt\left( \frac{\gamma}{4}+\frac{\lambda}{4\gamma^2} \right) ( 2\,\N + 1)
        + \frac{\lambda\N(\N-1)}{4\gamma^2} ,  \nonumber\\
    V(\gamma)=&&\hskip-18pt \left( \frac{\gamma}{4} - \frac{\lambda}{2\gamma^2} \right) (\Adag^2 + \A^2)
      \nonumber\\
    + \frac{\lambda}{4\gamma^2}\left( \frac{1}{6}\right.&&\hskip-18pt\left.(\Adag^4 + \A^4)
    + \frac{2}{3} (\Adag^3 \A + \Adag \A^3) \right) .\nonumber\\
\label{pertdecomplast}
\ea
Now, since I have a different perturbation theory defined for each choice of $\gamma$,
I need a principle for fixing $\gamma$.  This is
where the {\it adaptive\/} comes into adaptive perturbation theory.  The key notion is that I will
use a simple variational calculation, adapted to the eigenvalue to be calculated, in order to pick
that value of $\gamma$ that gives the most convergent perturbation expansion.

To set up this variational calculation I first define
a $\gamma$-dependent family of Fock-states, $\ket{\N}$.
The $\gamma$-dependent vacuum state,
$\ket{0_\gamma}$, is defined by the condition
\be
    \A \ket{0_\gamma} = 0 ,
\ee
and the $\gamma$-dependent $n$-particle state, {\it i.e.}, the state for which
\be
    N_\gamma \ket{\n} = n \ket{\n},
\ee
is just
\be
    \ket{\n} = \frac{1}{\sqrt{n!}} \,\Adag^n \ket{0_\gamma} .
\ee

The value of $\gamma$ used to define the adaptive perturbation
theory for the $n^{th}$ level of the anharmonic oscillator is determined by
requiring that it minimize the expectation value
\be
    E_n(\gamma) = \bra{\n} H \ket{\n} .
\label{engamma}
\ee
Eqs.~\ref{pertdecompone}-\ref{pertdecomplast} show that
this expectation value is equal to
\ba
    E_n(\gamma)\hskip -8pt &=& \hskip-8pt\bra{\n} H_0(\gamma) \ket{\n} = \left( \frac{\gamma}{4}
    + \frac{\lambda}{4\gamma^2} \right) ( 2\,n + 1) \nonumber\\
    &+& \frac{\lambda}{4\gamma^2} n\,(n-1).
\ea
Minimizing $E_n(\gamma)$ with respect to $\gamma$ gives
\be
    \gamma = \lambda^{1/3}\left( \frac{2\,(n^2 + n +1)}{2n+1} \right)^{1/3} .
\ee
At this point I substitute this value into Eq.~\ref{engamma} to obtain
\ba
    E_n(\gamma)_{\rm min}\hskip-8pt &=& \hskip-8pt \frac{3}{8}\,\lambda^{1/3} \left(2 n + 1 \right) ^{2/3}
    \left(2 n^2 + 2(n+1) \right)^{1/3}\nonumber\\
\ea
which, for large $n$, behaves as $\lambda^{1/3}\,n^{4/3}$ , which is the correct
answer.

The fact that all energies scale as $\lambda^{1/3}$ is an easily obtained exact
result and so, the non-trivial part of variational computation is the
derivation of the dependence of the energy on $n$. To see why all energies
are proportional to $\lambda^{1/3}$ it suffices to make the following
canonical transformation
\be
    x \rightarrow \frac{x}{\lambda^{1/6}} \quad ; \quad p \rightarrow \lambda^{1/6} p.
\ee
In terms of these operators, the Hamiltonian
of the pure anharmonic oscillator becomes
\vskip 20pt

\begin{table}[h]
\caption{A comparison of the zeroth order and second order perturbation results, for the
energy of the $n^{th}$ level of the pure anharmonic oscillator, to the exact answer for $\lambda = 1$
and widely varying values of $n$.}
\label{anharmtable}
\begin{center}
\begin{tabular}{|c|c|c|c|c|c|c|} \hline
$\lambda$ &  $n$ & Variational & $2^{nd}$ Order Perturbation & Exact & Variational \% Err & Perturbative \% Error \\
\hline
1.0 & 0  & 0.375 & 0.3712 & 0.3676 & 0.02 & 0.0098 \\
\hline
1.0 & 1  & 1.334   & 1.3195 & 1.3173   & 0.01   & 0.0017 \\
\hline
1.0 & 10 & 17.257  & 17.508 &  17.4228  & -0.009 & 0.0049 \\
\hline
1.0 & 40 & 104.317 & 105.888 & 105.360  & -0.009 & 0.0050 \\
\hline
\end{tabular}
\end{center}
\label{tableone}
\end{table}

\noindent
\be
    H = \lambda^{1/3} \left( \frac{1}{2} p^2 + \frac{1}{6} x^4 \right),
\ee
thus proving the claim.  A comparison of the variational computation and the
result of a second-order perturbation theory for $\lambda = 1$ and widely differing values of
$n$ is given in Table \ref{tableone}.  As advertised, we see that the
adaptive perturbation theory for each level converges rapidly, independent of
$\lambda$ and $n$.

Obviously, the same method can be used to study the case where $m^2 > 0$, except
that now one has to solve a cubic equation to determine $\gamma$ as a function
of $\lambda$,$m^2$ and $n$.  The general result that for each $n$, second order
perturbation theory is accurate to better than one percent still holds true.

\subsection{What does this have to do with quasi-particles?}

An interesting corollary to the adaptive perturbation theory
technique is that it provides an explicit realization of the
quasi-particle picture underlying much of many-body theory.  What we
have shown is that no matter how large the underlying coupling, the
physics of the states near a given $n_0$-particle state can be
accurately described in terms of perturbatively coupled eigenstates
of an appropriately chosen harmonic oscillator.  Of course, as we
have seen, the appropriately chosen harmonic oscillator picture
changes as $n_0$ changes.  Furthermore, the perturbatively coupled
states $\ket{n_{\gamma(n_0)}}$, for $n \approx n_0$, correspond to states
containing an infinite number of particles, if we choose as a basis
those states that correspond to a significantly different value of
$n_0$.

\newpage

\section{The Double Well}

The most general negative mass version of the anharmonic oscillator can,
up to an irrelevant constant, be written as:
\be
    H = \frac{1}{2}\, p^2 + \frac{1}{6}\,\lambda\, \left( x^2 - f^2 \right)^2 .
\label{doublewell}
\ee
Clearly, for a non-vanishing value of $f^2$, this potential has two minima
located at $x = \pm f$.  Thus, we would expect that the best
gaussian approximation to the ground state of this system can't be a
gaussian centered at the origin; classical intuition would imply that it
is a gaussian centered about another point, $x=c$.
In other words, if $\ket{0_\gamma}$ is a state centered
at $x=0$, it is better to adopt a trial state of the form
\be
    \ket{c,\gamma} = e^{-i c p}\,\ket{0_\gamma} .
\label{shiftedstate}
\ee

Since
\be
    e^{i c p}\, x\, e^{-i c p} = x + c ,
\ee
computing the expectation value of the Hamiltonian, Eq.~\ref{doublewell},
in the state specified in Eq.~\ref{shiftedstate}, is the same as computing the expectation
value of the Hamiltonian obtained by replacing the operator $x$ by $x+c$, in the state
$\ket{0_\gamma}$.  In the language of the previous sections, this is equivalent to introducing
the annihilation and creation operators $\Adag$ and $\A$ as follows:
\ba
    x &=& {1 \over \sqrt{2 \gamma}}\left( \Adag + \A \right) + c , \nonumber\\
    p &=& i\,\sqrt{\gamma \over 2} \left( \Adag - \A \right).
\label{shiftedops}
\ea
The expectation value of this Hamiltonian in the state $\ket{0_\gamma}$ is
\ba
    {\cal E}(c,\gamma) &=&\left( \frac{\gamma}{4} + \frac{\lambda f^4}{6} - \frac{\lambda f^2 c^2}{3}
    + \frac{\lambda c^4}{6} + \frac{\lambda c^2}{2 \gamma}\right.\nonumber\\
    &+& \left.\frac{\lambda}{8 \gamma^2}
    - \frac{\lambda f^2}{6 \gamma} \right) ,
\ea
which should be minimized with respect to both $\gamma$ and $c$ in order to define
the starting point of the adaptive perturbation theory computation.

\begin{figure}[ht]
\begin{center}
\leavevmode
\psfig{file=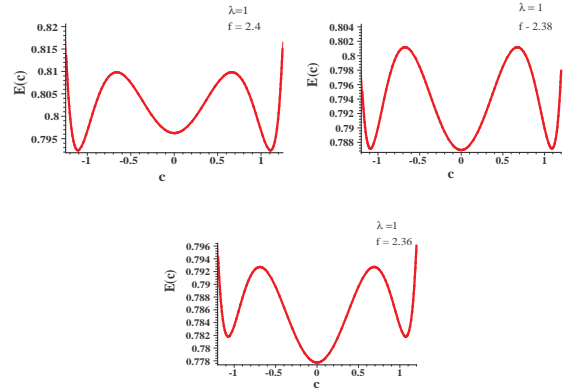,width=3in}
\end{center}
\vskip -30pt
\caption{Effective potential as a function of $c$, showing two degenerate local minima
at $c=\pm c_{\rm min}$ and one minimum at $c=0$. The three plots are for $\lambda=1$ and different
values of $f$ and show how the minimum at $c=0$ changes from lying higher than the minima at
$c=\pm c_{\rm min}$ to being the global minimum.  If this were field theory this would be
characteristic of a first order phase transition.}
\label{tricritone}
\end{figure}

To see what these equations tell us, it is convenient to hold $c$
fixed and solve for the value of $\gamma$ that minimizes ${\cal
E}(\gamma, c)$; call this $\gamma(c)$, and then plot ${\cal
E}(\gamma(c), c)$ for various values of $\lambda$ and $f^2$.  Three
such plots are shown in Fig.~\ref{tricritone}.  The first plot is
for a value of $f$ which is large enough so that the lowest energy
is obtained for a gaussian shifted either to the right or left by
the amount $c = \pm c_{\rm min}$.  While there is a local minimum at
$c=0$ it has a higher energy than the shifted states.  Things change
as one lowers the value of $f$.  Thus, the second plot shows that
for lower $f$ the shifted wavefunctions and the one centered at zero
are essentially degenerate in energy.  For a slightly smaller value
of $f$ things reverse and the unshifted wavefunction has a lower
energy than the shifted ones.
\vskip -20pt
\vskip -20pt
While, there would seem to be nothing wrong with the situation shown
in Fig.~\ref{tricritone}, it is problematic when one applies the same
sort of analysis to negative mass $\phi^4$ field theory in
$1+1$-dimensions.  In this case, if
one were to add a term like $J \phi$ to the Hamiltonian, this
result would imply the existence of a first order phase transition when $J$
reached some finite value.  At this point
the expectation value of $\phi$ in the ground-state would jump
discontinuously from zero to a non-zero value. It has been
rigorously shown that such a first order phase transition at a non-vanishing
value of $J$ cannot occur\cite{Simonetal,variational}.

\section{Doing Better}

\begin{figure}[h!]
\begin{center}
\leavevmode
\psfig{file=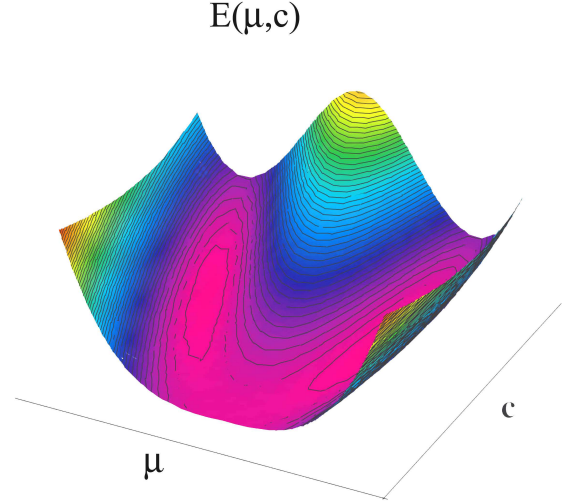,width=3in}
\end{center}
\vskip -20pt
\caption{Effective potential as a function o f $c$, one global minimum.}
\label{shiftedgaussone}
\end{figure}

Clearly, to do better, it is necessary to do something different.  The
solution is remarkably simple.  The trick is to take advantage of the fact
that $H$ contains a term linear in $\Adag$ and $\A$.
\ba
    H &=& \frac {\gamma}{4} - \frac {\lambda f^2}{6 \gamma} + \frac {\lambda f^4}{6} + \frac {\lambda}{8 \gamma^2}
    - \frac {\lambda f^2 c^2}{3} \nonumber\\
    + \hskip-10pt&&\hskip-14pt\frac {\lambda c^2}{2 \gamma}+ \frac {\lambda c^4}{6}
    + \left[ \frac {\gamma}{2} + \frac {\lambda}{2 \gamma^2} + \frac { \lambda ( 3\,c^2 - f^2)}{3 \gamma} \right]
    {\N} \nonumber\\
    +\hskip-10pt&&\hskip-14pt \frac {\lambda}{4 \gamma^2} \, {\N}({\N} - 1 ) \nonumber\\
    +\hskip-10pt&&\hskip-14pt
    \left[\frac{\sqrt{2}\lambda\,c\,(c^2-f^2)}{3\sqrt{\gamma}}\right]\left({\cal A}^{\dag} + {\cal A} \right)\nonumber\\
\ea
To do this introduce a trial state of the form.
\be
     \psi(\gamma,c,\alpha) =
     e^{ i c p} \left( \cos(\alpha)\, \ket{0_\gamma} + \sin(\alpha)\, \ket{1_\gamma} \right).
\ee
Varying over $\alpha$ is the same as minimizing the $2\times 2$
Hamiltonian of the zero and one particle states for fixed $\gamma$.
Then, having done this, minimize over $\gamma$ and $c$. The typical
result is seen in Fig.~\ref{shiftedgaussone}, where, as you can see,
the minimum at $c=0$ has all but disappeared.

\section{Large n}

For larger values of $n$, things change, even for $f \gg 0$.
Generically, what happens is that as $n$ grows $c_{\rm min}$ tends to zero, but doesn't quite
get there.  Thus, as for the case $n = 0$,
there are still two degenerate minima corresponding to equal and opposite values of $c_{\rm min}$
and so it is possible to lower the energy by forming the states
\ba
   \ket{\psi_{\rm even}} &=& e^{ i c p} \left( \cos(\alpha)\, \ket{n_\gamma} + \sin(\alpha)\, \ket{n+1_\gamma} \right)
   \nonumber\\
   &+& e^{ -i c p} \left( \cos(\alpha)\, \ket{n_\gamma} - \sin(\alpha)\, \ket{n+1_\gamma} \right)\nonumber\\
\ea
and
\ba
   \ket{\psi_{\rm odd}} &=& e^{ i c p} \left( \cos(\alpha)\, \ket{n_\gamma} + \sin(\alpha)\, \ket{n+1_\gamma} \right)
   \nonumber\\
   &-& e^{ -i c p} \left( \cos(\alpha)\, \ket{n_\gamma} - \sin(\alpha)\, \ket{n+1_\gamma} \right) .\nonumber\\
\ea
The result of such a computation is to show that as $n$ grows the splitting between these states
grows. Eventually, no matter what value is assigned to $f$ (so long as it is finite) there
is a value of $n$ for which the splitting between the states $\ket{\psi_{\rm even}}$
and $\ket{\psi_{\rm odd}}$ becomes of order unity.  This is the point at which it makes
no sense to talk about tunneling between states defined on one or the other side of the
potential barrier.

\section{Applying It To Field Theory}

For the purpose of this talk I will limit my discussion to the case of $\phi^4$-field theory whose Hamiltonian
is
\ba
H\hskip-10pt&=&\hskip-10pt\int d^nx \left[ \frac{1}{2} \Pi_{\phi}(x)^2 + \frac{1}{2} \left(\nabla \phi(x)\right)^2
               + \frac{\lambda}{6} \phi(x)^4 \right] .\nonumber\\
\ea
It is customary to rewrite the operators $\phi(x)$ and $\Pi_{\phi}(x)$ in terms
of their Fourier transforms; {\it i.e.\/}
\ba
    \phi(\vec{k}) &=& \frac{1}{\sqrt{V}} \int d^nx e^{-i \vec{k}\cdot\vec{x} } \phi(x), \nonumber\\
    \Pi_{\phi}(\vec{k}) &=& \frac{1}{\sqrt{V}} \int d^nx e^{-i \vec{k}\cdot\vec{x} } \Pi_{\phi}(x),
\ea
where $V$ stands for the volume of the system.  (I have assumed the system is in a finite volume so that the
momenta become discrete and the operators become well defined.)

This Hamiltonian can be rewritten in terms of these momentum space operators as
\ba
    H =  \sum_{\vec{k}} &&\hskip-24pt \left[ \frac{\Pi_\phi(-\vec{k}) \Pi_\phi(\vec{k})}{2}
    +\frac {({\vec{k}}^2 + m^2)}{2}\,\phi(\vec{k})\,\phi(-\vec{k})\right]\nonumber\\
    + \frac{\lambda}{6} && \hskip-18pt \frac{1}{V} \sum_{\vec{k}_i} \phi(\vec{k}_1)\,\phi(\vec{k}_2)
\phi(\vec{k}_3)\,\phi(\vec{k}_4) \delta(\sum_{i=1}^4 \vec{k}_i). \nonumber\\
\ea
To define the adaptive perturbation theory calculation I introduce
$\gamma(k)$-dependent annihilation and creations operators as follows,
\ba
    \phi(\vec{k}) &=& i\,\sqrt{\frac{\gamma(\vec{k})}{2}} ( {\cal A}(-\vec{k})^{\dag} - {\cal A}(\vec{k}) )
    \nonumber\\
    \Pi(\vec{k}) &=& \frac{( {\cal A}(-\vec{k})^{\dag} + {\cal A}(\vec{k}) )}{\sqrt{2\,\gamma(\vec{k})}},
\ea
The {\it vacuum\/} state associated with this choice of $\gamma(\vec{k})$'s,
\be
    \ket{vac} = \prod_k \ket{0_{\gamma(k)}},
\ee
is defined by the condition that it be annihilated by all the ${\cal A}(\vec{k})$'s.

As in the simpler example I determine the $\gamma(k)$'s by
minimizing the vacuum expectation value of the
Hamiltonian in this trial state.  It follows directly from these definitions that the function to be
minimized is
\ba
\bra{vac}\, H\, \ket{vac}\hskip-10pt&=&\hskip-10pt  \sum_k \left[ \frac{\gamma(\vec{k})}{4}
+ \frac{( \vec{k}\cdot\vec{k} + m^2 )}{4\,\gamma(\vec{k})} \right] \nonumber\\
+ \frac{\lambda}{4 V} &&\hskip-18pt\left[   \sum_{\vec{k}} \frac{1}{\gamma(\vec{k})} \right]^2.
\label{vacenergy}
\ea

Obviously, if the range of the momenta appearing in these sums is
unrestricted, these expressions diverge.
It is customary to deal with this problem, in the context of
ordinary perturbation theory, by regulating the integrals and adding
counterterms to the Lagrangian to cancel divergences.  Since I
wish to discuss this theory non-perturbatively, I will adopt a
different strategy.  I will render the theory well defined by
restricting the operators $\phi(\vec{k})$ and $\Pi_\phi(\vec{k})$ to
be finite in number.  This can be accomplished in a variety of ways,
but all amount to restricting the range of $\vec{k}$'s which appear
in the Fourier transform.

Minimizing Eq.~\ref{vacenergy} with respect to each $\gamma(\vec{k})$ yields
\be
    \gamma(k)^2 = \vec{k}\cdot\vec{k} + m^2
    + 2\,\lambda\,\left[\frac{1}{V} \sum_{\vec{k}'} \frac{1}{\gamma(\vec{k}')} \right].
\label{generick}
\ee
In particular, the equation for $\vec{k}=0$ is
\be
    \gamma(0)^2 = m^2 + 2\,\lambda\,\left[ \frac{1}{V} \sum_{\vec{k}'} \frac{1}{\gamma(\vec{k}')} \right] ,
\ee
which can be substituted into Eq.~\ref{generick} to give
\be
    \gamma(\vec{k})^2 = \vec{k}\cdot\vec{k} + \gamma(0)^2 .
\ee
If we use this to rewrite the equation for
$\gamma(0)$, it becomes the non-perturbative equation
\be
    \gamma(0)^2 = m^2
    + 2\,\lambda\,\frac{1}{V} \sum_{\vec{k}'} \frac{1}{\sqrt{(\vec{k}')^2 + \gamma(0)^2}}.
\ee
Taking $V$ to infinity and converting the sum over $\vec{k}$ to an integral,
we obtain an equation which is reminiscent of the Nambu Jona-Lasinio equation.
\ba
    \gamma(0)^2-m^2\hskip-7pt&=&\hskip-7pt
     \frac{(2\lambda) 2^p}{(2\pi)^p}\int_0^{\Lambda} k^{p-1} dk
    \frac{1}{\sqrt{\vec{k}\cdot\vec{k} + \gamma(0)^2}}. \nonumber\\
\label{NJL}
\ea
In fact Eq.~\ref{NJL} should not be thought of as an equation
for $\gamma(0)$, but rather as an equation for $m^2$; {\it i.e.}, it
should be rewritten as
\ba
    m^2 = \gamma(0)^2&&\hskip-20pt
    -\frac{ (2\lambda)2^p}{(2\pi)^p} \int_0^{\Lambda} k^{p-1} dk
    \frac{1}{\sqrt{\vec{k}\cdot\vec{k} + \gamma(0)^2}} .\nonumber\\
\ea
This shows that for any arbitrarily chosen value of $\gamma(0)$, this equation
determines the value of $m^2$ for which the chosen value of $\gamma(0)$
will minimize the ground state energy density.
This is, of course, nothing but a non-perturbative way of determining
the leading mass renormalization counter term.

\section{Wavefunction Renormalization}

To this point I have shown how a simple variational calculation
captures the general notion of mass renormalization in a non-perturbative
manner.  I now wish to briefly describe what one has to do to
capture wavefunction and coupling constant renormalization.
The trick is to proceed as in the double well problem
and generalize the trial state to include the effects of
the ${{\cal A}^{\dag}}^4$ and ${\cal A}^4$ terms which appear
in the normal ordered Hamiltonian; {\it i.e.} terms of the form
\ba
\sum_{k_1,k_2,k_3}&&\hskip-20pt\frac {1}{4\,\sqrt{\gamma(k1)\,\gamma(k_2)\,
\gamma(k_3)\,\gamma(k_4)}}\times\nonumber\\
&& \left[ \AAdag_{k_1}\,\AAdag_{k_2}\,\AAdag_{k_3}\,\AAdag_{k_4}
+  \AA_{k_1}\,\AA_{k_2}\,\AA_{k_3}\,\AA_{k_4}\right] \nonumber\\
\label{justfour}
\ea
To allow these terms contribute to the ground state energy
I have to add to the vacuum state a general four particle state;
{\it i.e.}, I need a trial state of the general form
\be
    \ket{vac} + \sum \alpha(k_1,k_2,k_3) \ket{4-particle},
\ee
where the sum is assumed to go over all possible four-particle
states. The question is "How do we minimize over the
$\alpha(k_1,k_2,k_3)$'s?".

Actually this question can be finessed since it is equivalent
to finding the ground state energy density of a system
where the Hamiltonian is truncated to the vacuum state and all
possible four particle states.  While it is hard to do this for the
full Hamiltonian, it can be done for the
case where the full Hamiltonian is limited to the part
indicated in Eq.~\ref{justfour}.
In this case the general solution, which will resum
the important non-perturbative effects in $\lambda$, can
be obtained by considering
the resolvent operator.
Dividing $H$ into the part diagonal in
the number operator and the terms in Eq.~\ref{justfour}; {\it i.e.},
\be
    H = H_0 + \lambda\,V ,
\ee
it is easy to show that the resolvent operator satisfies the integral equation
\be
 \frac {1}{ H - z } = \frac{1}{H_0 - z}- \frac{\lambda}{H_0 - z}V\frac{1}{H - z}.
\ee
This equation is customarily solved by iteration to give the following series;
\ba
    \frac {1}{ H - z } = \frac{1}{H_0 - z}&&\hskip-20pt - \frac{\lambda}{H_0 - z}V\frac{1}{H_0 - z} \nonumber\\
 + \frac{\lambda^2}{H_0 - z} V \frac{1}{H_0 - z}&&\hskip-20pt V \frac{1}{H_0 - z} \nonumber\\
 - \frac{\lambda^3}{H_0 - z} V \frac{1}{H_0 - z}&&\hskip-20pt V \frac{1}{H_0 - z} V  \frac{1}{H_0 - z} \nonumber\\
 + \frac{\lambda^4}{H_0 - z} V \frac{1}{H_0 - z}&&\hskip-20ptV \frac{1}{H_0 - z} V
 \frac{1}{H_0 - z}V\frac{1}{H_0 - z} \nonumber\\
  + \ldots\ldots \phantom{\frac{\lambda^4}{H_0 - z}\, V} && \qquad
\ea
Since $V$ only links the vacuum to the four particle states and then links four
particle states back to the vacuum, this series simplifies to
\ba
 \bra{0} \frac {1}{ H - z } \ket{0}&=&\frac{1}{E_0 - z} \nonumber\\
 + \frac{1}{E_0 - z}\hskip-10pt&&\hskip-18pt\V \nonumber\\
 + \left(\frac{1}{E_0 - z}\right.\hskip-10pt&&\hskip-18pt\left.\V\right)^2 + \ldots \nonumber\\
= \frac{1}{E_0-z}\hskip-10pt&&\hskip-18pt\left( \sum_{j=0}^{\infty} \V \right)^j \nonumber\\
= \phantom{\left(\frac{1}{E_0 - z}\right.}\hskip-20pt&&\hskip-34pt\frac{1}{E_0 - z - \Vee}.\nonumber\\
\ea
Clearly, since the poles of the resolvent operator correspond to the eigenvalues
of the Hamiltonian it is only necessary to find that zero of the function
\be
    E_0 - z - \Vee = 0
\label{zstart}
\ee
which lies to the value $E_0$ in the limit $\lambda \rightarrow 0$.
While this looks like a difficult problem it can be solved to arbitrary
accuracy by converting the problem into an equation which can be solved
iteratively.  To show this I will begin by defining $\delta z$ as
\be
   z = E_0 - \lambda \sqrt{ \sum_n \vert V \vert_{0n}^2}\ \delta z,
\ee
and then use this definition to rewrite Eq.~\ref{zstart} as an integral equation for $\delta z$; {\it i.e.},
\be
    \delta z = \sum_n \frac {\vert \bar{V} \vert_{0n}^2}{\frac {E_n-E_0}{\lambda} + \delta z },
\ee
where I have defined $ \bar{V}_{0n}$ to be
\be
 \bar{V}_{0n} = \frac {\vert \bra{0} V \ket{n} \vert}{\sqrt{ \sum_j \vert V \vert_{0j}^2 }}.
\ee
It is clear that with these definitions, so long as the range of $E_n-E_0$ is bounded, in the limit
$\lambda \rightarrow \infty$ the solution to the integral equation is
$\delta z = 1$.  In other words, in the limit of large $\lambda$ the energy shift
is proportional to $\lambda$ instead of the perturbative behaviour which goes like $\lambda^2$.

\begin{figure}[ht!]
\begin{center}
\leavevmode
\psfig{file=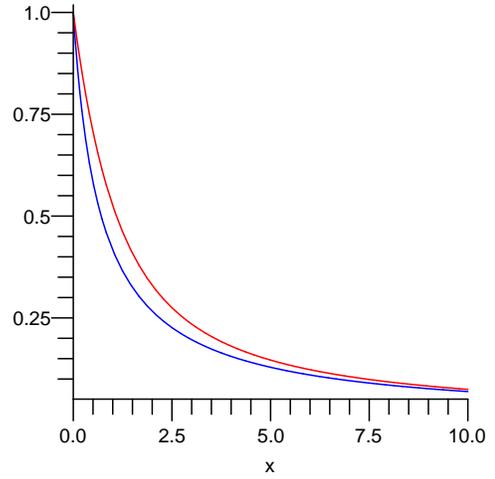,width=3in}
\end{center}
\vskip -20pt
\caption{Comparison of exact energy as a function of $x$ (red curve) compared to $\delta z_1$ (blue curve)}
\label{integone}
\end{figure}

Given the large $\lambda$ value of $\delta z$ it is possible to
give an iteration procedure for solving for $\delta z$ for arbitrary values of $\lambda$.
This is done by defining a sequence of values $z_j$ by the recursion relation
\ba
    \delta z_0 &=& 1 \nonumber\\
    \delta z_1 &=& \sum_n \frac{\vert \bar{V} \vert_{0n}^2}{\frac{E_n-E_0}{\lambda} + \delta z_0 } \nonumber\\
    \delta z_n &=& \sum_n \frac{\vert \bar{V} \vert_{0n}^2}{\frac{E_n-E_0}{\lambda} + \delta z_{n-1}},
\ea
The desired value of $\delta z$ is obtained by taking the limit $n \rightarrow \infty$.
As an example, consider the case in which $H$ is a $2\times 2$ matrix with two arbitrary
entries on the diagonal $E_0$ and $E_1$; {\it i.e.},
\be
  H = \left( \begin{array}{*{2}{c@{\quad}}}
    E_0  & x \nonumber\\
     & \\
    x & E_1\nonumber\\
    \end{array}
    \right).
\ee
Since it is trivial to diagonalize this matrix it is easy to compare, for the case $E_0=1$ and $E_1=2$,
the iterative solution for arbitrary $x$ to the exact answer.
Figure~\ref{integone} shows how a single
iteration manages to reproduce the behavior of the exact answer to pretty good accuracy over the whole range
of $x$.

\vskip -30pt
\begin{figure}[h!]
\begin{center}
\leavevmode
\psfig{file=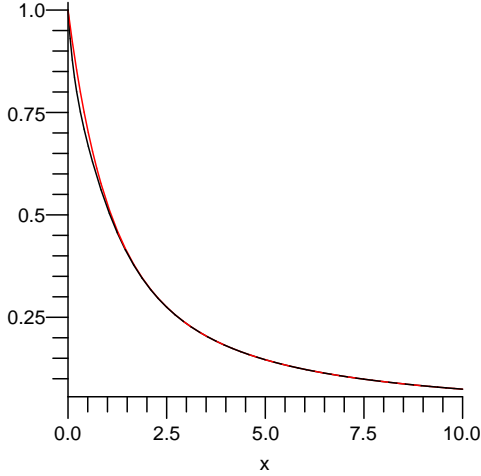,width=3in}
\end{center}
\vskip -20pt
\caption{Comparison of exact energy as a function of $x$ (red curve) compared to $\delta z_3$ (blue curve)}
\label{integthree}
\end{figure}

After three iterations it is much harder to distinguish the curves, see Fig.~\ref{integthree}.
Doing more iterations just produces better and better accuracy over the entire range.  By twelve iterations
one is accurate to very high accuracy except in a very small band around $x \approx 0.8$, where the fractional
error is $\approx 0.015$.  The same computation can be done for the more realistic case where $V$ connects
us to a large number of momenta, with similar results.

Having established that it is in principle possible to obtain the shift in the ground state energy
due to the part of the Hamiltonian which mixes the vacuum with four particle states,
it is time to go back to our discussion of the adaptive perturbation theory scheme.
Given the preceding discussion it is easy to show that differentiating the expectation value
of the Hamiltonian in our trial state will yield a set of equations for the $\gamma(k)$'s of the
general form
\ba
\frac{1}{4} - \frac{( k^2 + m^2 ) }{4\,\gamma(k)^2} \hskip-20pt&&
- \frac {\lambda}{2 \gamma(k)^2} \left[\frac{1}{V} \sum_{k'} \frac {1}{\gamma(k')} \right] \nonumber\\
- \frac {\lambda^2}{\gamma(k)^2}\hskip-10pt&&\hskip-10pt\,\Sigma(k^2,\lambda,\gamma(k_i)) = 0,
\ea
where the function $\Sigma(k^2,\lambda,\gamma(k_i))$ is obtained by differentiating the
solution for $\delta z$ obtained from our integral equation.  In writing things in this
form I used the fact, which can be obtained from the integral equation,  that
differentiating $\delta z$ with respect to a given $\gamma(k)$ produces a factor
of $\gamma(k)^{-2}$ multiplied by a function of $k^2$,$\lambda$ and all the $\gamma(k_i)$.
As in the earlier discussion of mass renormalization, it is convenient to rewrite these equations
as
\ba
    \gamma(k)^2 = k^2 + m^2 &+& 2\,\lambda\,\left[\frac{1}{V} \sum_{k'} \frac{1}{\gamma(k')} \right]\nonumber\\
&+& \lambda^2\,\Sigma(k^2,\lambda,\gamma(k_i))
\ea
and then observe that this means that
\ba
    \gamma(0)^2 = m^2  &+& 2\,\lambda\,\left[ \frac{1}{V} \sum_k \frac{1}{\gamma(k)} \right]\nonumber\\
                &+& \lambda^2\Sigma(0,\lambda,\gamma(k_i)).
\ea
Substituting this into the equation for a generic value of $k$ yields
\ba
    \gamma(k)^2 &=& k^2 + \gamma(0)^2 \nonumber\\
    &+& \lambda^2 \left( \Sigma(k^2,\lambda,\gamma(k_i))
- \Sigma(0,\lambda,\gamma(k_i)) \right).\nonumber\\
\label{wvrenone}
\ea

At this point the role of wavefunction renormalization becomes obvious.  If, as is conventional, we
Taylor series expand $\Sigma(k^2,\lambda,\gamma(k_i))$ so as to rewrite Eq.~\ref{wvrenone} as
\ba
    \gamma(k)^2 &=& (1 + \frac {\partial}{\partial k^2} \Sigma(k^2),\lambda,\gamma(k_i))k^2
    + \gamma(0)^2 \nonumber\\
    &+&\lambda^2 \left( \Sigma(k^2,\lambda,\gamma(k_i))
- \Sigma(0,\lambda,\gamma(k_i)) \right.\nonumber\\
& -& \left.k^2\frac{\partial}{\partial k^2} \Sigma(0,\lambda,\gamma(k_i))\right).
\ea
This can be put into a conventional form
\be
    \gamma(k)^2 = k^2 + \gamma(0)^2 + \lambda^2 \bar{\Sigma}(k^2,\lambda,\gamma(k_i))
\ee
by introducing a rescaling of the fields and the overall Hamiltonian by
\ba
    \phi &\rightarrow& Z_\phi^{1/2} \phi  \quad ; \quad
    \Pi_\phi \rightarrow \frac {\Pi_\phi}{Z_\phi^{1/2}} \nonumber\\
    H &\rightarrow& Z_\phi H .
\ea
Thus, we see that wavefunction renormalization is just that, an overall change in the
$\gamma(k)$'s.  Note that the rescaling of the Hamiltonian is necessary because we are
working in a Hamiltonian and not a manifestly covariant formalism.

Finally, I will just say a few words about coupling constant renormalization and why
$\phi^4$ theory doesn't exist in four dimenstions, at least if one attempts to remove
the cutoff.  Clearly, once the $\gamma(k)$'s have been chosen, the only remaining issue
is determining those values of $\lambda$ for which some physical quantity come out finite.
For the purposes of this discussion I can choose this to be the energy of a trial state
containing only two particles of momentum zero.

As in earlier discussions I will restrict this analysis to the effect of terms
in $H$ which take two particles to two particles.  In this case I can employ a similar argument,
based upon studying the matrix element of the resolvent operator in this
zero-momentum state, to show that in four dimensions this energy
is given by a series in $\lambda \left(\lambda \ln(\Lambda) \right)^n$.
From this it follows that the only way to have this come out finite is to take
$\lambda \rightarrow 1/\ln(\Lambda)$.  But this, of course, implies that there is no interaction.

\section{The Spectrum Isn't Boost Invariant}

The final point which should be touched upon before closing my talk is
the observation that at first glance an equation of the form
\be
    \gamma(k)^2 = k^2 + \gamma(0)^2 + \lambda^2 \bar{\Sigma}(k^2,\lambda,\gamma(k_i))
\ee
would seem to be problematic, since $\gamma(k)$ for $k \ne 0$ is not just
related by a boost to the value for $k=0$.  In fact, this is not a bug,
it is a feature.  What it says is that while it is possible to choose the
$\gamma(k)$'s to give a good perturbation perturbation theory, with this
choice of $\gamma(k)$'s the operators ${\cal A}^{\dag}(k)$
do not create a true asymptotic states which should
be used to compute scattering amplitudes.  Even if I assume that the
state created by ${\cal A}^{\dag}(0)$ is a good approximation to the zero
momentum state, the state for $k \ne 0$ should be obtained by applying
the boost operator to this state.  For an interacting theory this is
certainly a multi-particle state.  Thus, it follows that adopting the
formalism of adaptive perturbation theory forces implies that the scattering
problem must be handled as in the
parton picture.  In other words, the computation of a scattering process
intrinsically has two parts; first, it is necessary to find the
parton wavefunction for the asymptotic scattering states and then,
this wavefunction, together with the explicit form of
the Hamiltonian written in terms of the annihilation and creation operators
associated with the specific choice of $\gamma(k)$'s, should be used to
compute scattering amplitudes.  This point, together with the observation that
applying a boost (written in terms of the relevant annhilation and creation
operators) to a given state implies a kind of Alterelli-Parisi equation,
requires much more thought.

\section{Summary}

In this talk I began by showing you how to convert problems, which
in the past were thought to be impossible to deal with perturbatively,
can be easily done using the method I have called adaptive perturbation
theory.  I then went on to show how these methods can be extended to
give a very pretty picture of the structure of renormalization
in a non-perturbative context.  These ideas can be extended
to a theory which includes fermions by adding adding a general Bogoliubov
transformation for the fermion fields defined in momentum space.  A bit
more work has to be done to extend the tricks to cover the case of
compact Abelian gauge-fields.  However, a paper by David Horn and
myself\cite{Horn:1981ht}, extended in the obvious way to fit it into
the framework of adaptive perturbation theory, showed how to do this for
the case of compact QED.  This paper showed that the method is quite capable of
extracting the interesting non-perturbative structure of confinement in
this theory in any number of space-time dimensions.  While the path
towards extending these ideas to the case of a non-abelian gauge theory is
not so obvious, I nevertheless believe that it is possible.

\end{document}